\documentclass[preprint,showpacs,prb]{revtex4}

\usepackage{graphicx}

\begin{document}

\preprint{}

\title{Plasmon-Polariton Waves in Nanofilms on One-Dimensional 
Photonic Crystal Surfaces }

\author{Valery N. Konopsky}
 \email{konopsky@gmail.com}
 \homepage{http://www.isan.troitsk.ru/dsss/lsss/kvn}

\affiliation{
Institute of Spectroscopy,
Russian Academy of Sciences,
     Troitsk, Moscow region, 142190, Russia\-.}

\date{\today}

\begin{abstract}

The propagation of bound optical waves along the surface of 
a one-dimensional (1\mbox{-}D) photonic crystal (PC) structure is considered.
 A unified description of the waves in 1\mbox{-}D~PCs for both s- and p-polarizations is done via 
an impedance approach.
 A general dispersion relation that is valid for optical surface waves with both polarizations
is obtained, and conditions are presented for long-range propagation of  plasmon-polariton
waves in nanofilms (including lossy ones) deposited on the top of the 1\mbox{-}D~PC
structure.
 A method   is described for designing  1\mbox{-}D~PC structures to fulfill the conditions required for the existence
of the surface mode  with a particular wavevector at a particular wavelength.
 It is shown that the propagation length of the long-range surface plasmon-polaritons 
can be maximized by wavelength tuning, which introduces a slight asymmetry in the system.

\end{abstract}


\maketitle

\section{Introduction}
Optical surface waves  (SWs) are excitations of electromagnetic (EM)
modes that are  bound to the
interface between two media.
 The maximum  EM  field
strength of the SWs is located near the interface, 
so these waves have a  high sensitivity  to the surface condition. 
 The SWs that have been employed in  the widest array of applications are 
surface plasmon polariton (SPP) waves, which are p-polarized optical
surface waves propagated along a metal-dielectric
interface~\cite{Raether1988}.  
 The sensitivity of SPPs  has been used in many surface plasmon
resonance (SPR) applications in which
a shift of an SPR dip is measured, from
biosensors used for detecting  biomolecules in a liquid,
to gas sensors for detecting  trace impurities  in the
air~\cite{Liedberg1983}.  

A limiting factor for   SPR  sensitivity
in these applications is the  limited propagation length of
SPPs  due to a strong intrinsic damping of their EM field  in
metal.  
 Even when  the ``best plasmonic'' metals, such as silver
and gold,  are used, the SPP propagation length is only about ten micrometers
in the optical frequency range.  
 Other metals  do
not practically support any SPP propagation at visible frequencies.
 One way to increase the SPP propagation length and, consequently,
the SPR sensitivity is to use  long-range SPPs (LRSPPs),
which can be achieved using a thin metal film embedded between two
dielectrics with identical refractive indices
(RIs)~\cite{Sarid1981, Sarid1983, Sambles1991}.  

In the work~\cite{PRL2006},  another method for the excitation of LRSPPs was reported,
in which the thin
metal film was embedded between the  medium being studied (with any RI, even a gas medium) and 
the 1\mbox{-}D photonic crystal.
 Photonic crystals (PCs) are  materials that possess a periodic modulation of their
refraction indices on the scale of the
wavelength of light~\cite{Yablonovitch1993}.
 Such materials can exhibit photonic band gaps that are very much like the
electronic band gaps for electron waves traveling in the
periodic potential of the crystal.
 In both cases, frequency intervals exist in which wave propagation is
forbidden.
This analogy may be extended~\cite{Kossel1966} to include surface
levels, which can exist in band gaps of electronic crystals.
 In PCs, they correspond to optical surface waves with dispersion curves
located inside the photonic band gap.

\begin{figure}[h]
	\centering
\includegraphics{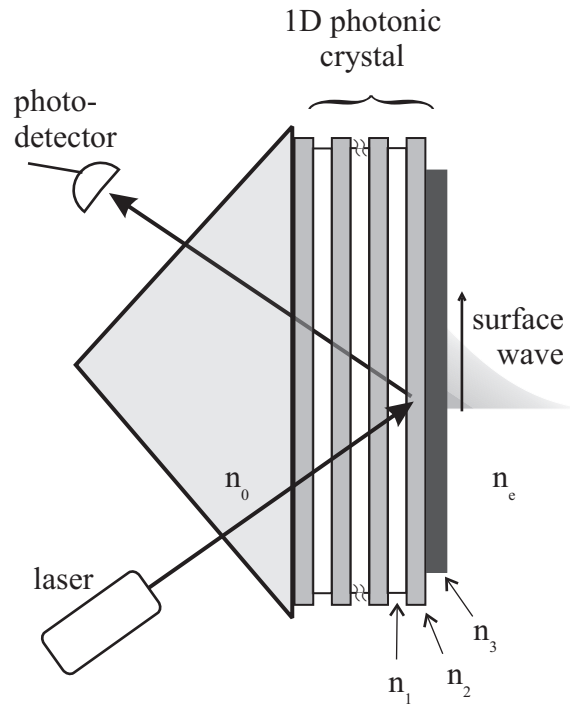}
		\caption{Excitation of optical surface waves in a Kretschmann-like 
		 scheme.}
	\label{setup}
\end{figure}

The one-dimensional photonic crystal (1\mbox{-}D~PC) is a simple periodic multi-layer stack.
 Optical surface modes in 1\mbox{-}D~PCs were studied in the 1970s, both
theoretically~\cite{Yariv1977} and experimentally~\cite{Yariv1978}.
 Twenty years later, the excitation of optical surface waves in a
Kretschmann-like configuration  was demonstrated~\cite{Robertson1999a}.
 A scheme of the Kretschmann-like excitation of PC~SWs is presented in 
Fig.~\ref{setup}.
 In recent years, the PC~SWs have been used in ever-widening applications in the field
of optical sensors~\cite{AC2007,Robertson2005,Humana2009,BB2010,OL2009,NJP2009}.
 In contrast to SPPs, both p-polarized and  s-polarized optical surface waves 
(with a dielectric final layer of 
the 1\mbox{-}D~PC)~\cite{AC2007,Humana2009,BB2010}
can be used in   PC~SW sensor
applications. 
 However,  in applications  in which the LRSPP propagates in a metal nanofilm deposited
on an appropriate 1\mbox{-}D~PC surface (e.g., in  hydrogen 
detection~\cite{OL2009,NJP2009}),
only p-polarized waves can be  used.
 Therefore, a unified theoretical description of PC~SWs for both polarizations would be very useful,
and  such a description is presented in this article.
 Additionally, the current contribution 
describes conditions 
for the   propagation of LRSPP in  metal nanofilms (including lossy ones) that are
deposited on the top of the 
1\mbox{-}D~PC structure and, thereby, 
provides the theoretical background for the 
works~\cite{PRL2006,OL2009,NJP2009}.

\section{Dispersion relation for s- and p-polarized surface waves in impedance terms}

\subsection{Impedance approach: basic definitions and recursion relation for a multilayer}
 
The ``characteristic impedance'' of an optical medium is the ratio of the electric field amplitude to the magnetic field amplitude in this medium, i.e., $Z_\mathrm{char}=E/H=1/n$.
 The concept of impedance in the optics of homogeneous layers is based on a mathematical 
analogy between a cascade of transmission line sections and a multilayer optical coating.

\begin{figure}[h]
	\centering
\includegraphics{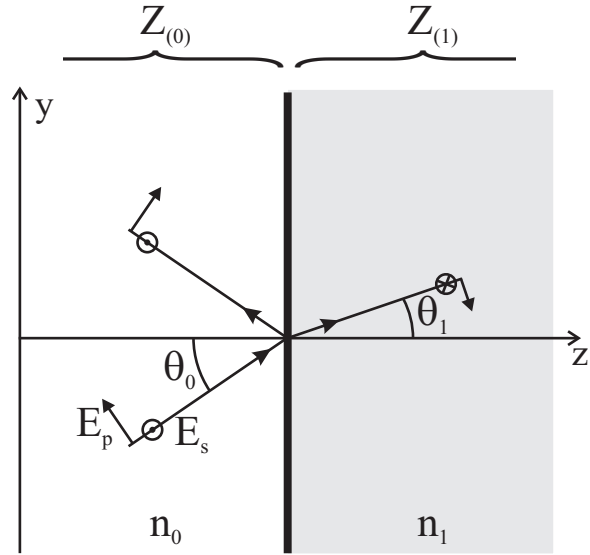}
		\caption{Reflection and transmission from a single interface.}
	\label{single}
\end{figure}

 For reflection from a plane interface, the useful value is the ``normal
impedance''~\cite{Brekhovskikh, Delano} $Z$, which is the ratio of the tangential components of the electric field to the  magnetic field:
\begin{equation}
Z=\frac{E_\mathrm{tan}}{H_\mathrm{tan}} \; .
\label{Z}
\end{equation}
Impedances for the s-polarized  wave (in which the electric field vector is orthogonal to the incident plane
--- TE wave) and for the p-polarized wave
(in which the electric field vector is parallel to the incident plane
--- TM wave)   are correspondingly:
\begin{eqnarray}
Z_s=&\displaystyle\frac{E_x}{H_y}
=&\displaystyle\frac{1}{n \cos(\theta)}   \qquad  \mbox{\small(for TE wave)}
\label{Zs}\\[8pt]
Z_p=&\displaystyle\frac{E_y}{H_x}
=&\displaystyle\frac{\cos(\theta)}{n}   \qquad  \mbox{\small(for TM wave)} \, .
\label{Zp}
\end{eqnarray}

For the interface between semi-infinite media $0$ and $1$ (Figure~{\ref{single}}), Fresnel's formula for reflection coefficients has a very simple form in the impedance terms:
\begin{equation}
R=\displaystyle\frac{
{Z}_{(1)}-{Z}_{(0)}}{
{Z}_{(1)}+{Z}_{(0)}}\, , 
\label{Rsp} 
\end{equation}
where $Z_{(j)}$ is the normal impedance of medium $j$, given by~(\ref{Zs}) or~(\ref{Zp}).
 Hereafter, the use of $R$ and $Z$ (without subscripts $s$ or $p$) means that the equation holds for 
both polarizations when the corresponding impedances $Z_s$ or $Z_p$ are inserted. 
 Fresnel's formulas for transmission coefficients are as follows:
\begin{eqnarray}
T_s&=&-2\displaystyle\frac{
Z_{s(1)}}{
Z_{s(1)}+Z_{s(0)}}   \\[8pt]
T_p&=&-2\displaystyle\frac{ n_0}{\displaystyle n_1}\displaystyle\frac{
Z_{p(0)}}{
Z_{p(1)}+Z_{p(0)}} \, . 
\label{Tsp}
\end{eqnarray}

In Figure~{\ref{single}}, the labeling of the s-polarized wave in the first and the second media by $\odot$ and $\otimes$ indicates the direction of the  electric field vector $\vec E$ after transmission/reflection in accordance with  our ``rule of signs.''

\begin{figure}[h]
	\centering
\includegraphics{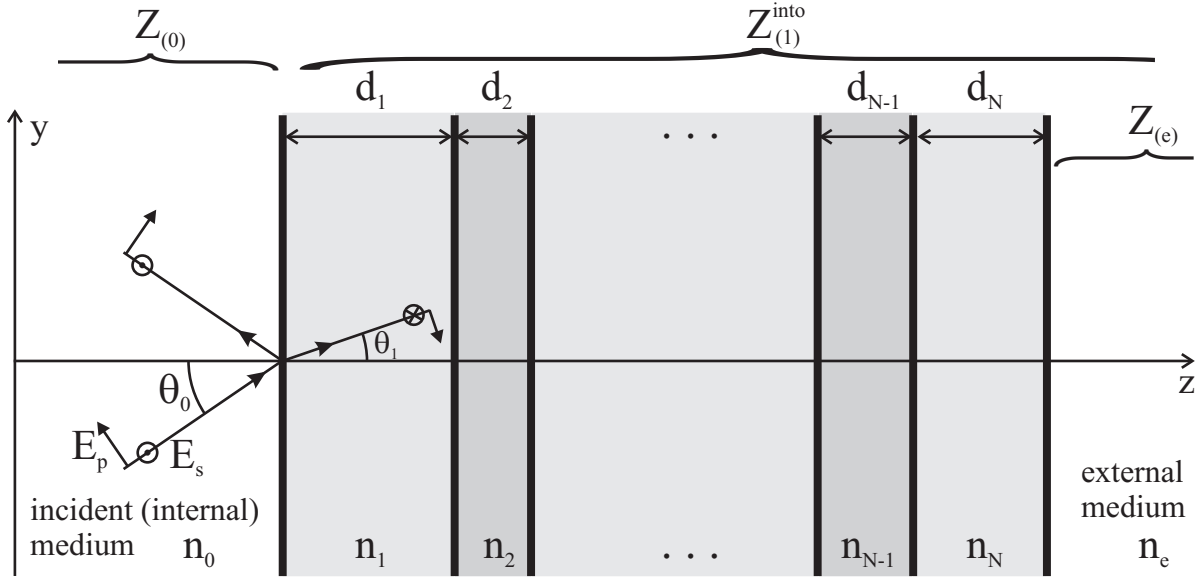}
		\caption{Reflection and transmission for a multilayer with N layers.}
	\label{multi}
\end{figure}

The equations for reflection coefficients of s- and p-polarized waves from any complex multilayer (see Figure~{\ref{multi}})  also have a form similar to equation~(\ref{Rsp}):
\begin{equation}
R=\displaystyle\frac{
Z_{(1)}^\mathrm{into}-{Z}_{(0)}}{
Z_{(1)}^\mathrm{into}+{Z}_{(0)}} \, , 
\label{R_into} 
\end{equation}
where $Z_{(1)}^\mathrm{into}$ is an 
apparent input impedance for a multilayer,
i.e., it is the impedance that is seen by an incoming wave as it approaches  to the interface.

If the multilayer is made up of  $N$ plane-parallel, homogeneous, isotropic dielectric layers (with refractive indices $n_j$ and geometrical thicknesses $d_j$, where $j=1,2, \ldots, N$) between semi-infinite incident~$_{(0)}$   and external~$_{(e)}$ media (see~Fig(\ref{multi})), the apparent input impedance $Z^\mathrm{into}_{(j)}$ of a semi-infinite external  medium~$_{(e)}$ and layers from $N$ to $j$  may be calculated by the following {\sl recursion relation}~\cite{Brekhovskikh, Delano}:
\begin{equation}
Z_{(j)}^\mathrm{into}=Z_{(j)}{\frac {{ Z^\mathrm{into}_{(j+1)}}-iZ_{(j)}\tan(\alpha_{j})}{Z_{(j)}-i{ Z^\mathrm{into}_{(j+1)}}\tan(\alpha_{j})}}
\label{Z_into}
\end{equation}
where $\alpha_j=k_{z(j)}\,d_j=(2\pi/\lambda)n_j\cos(\theta_j)\,d_j$; $j=N, N-1,\ldots, 2,1$  and 
$Z_{(N+1)}^\mathrm{into}=Z_{(N+1)}=Z_{(e)}$, $n_{N+1}=n_{e}$ while $d_{N+1}=d_{e}=0$ by definition.

Fresnel's formulas for  multilayer transmission coefficients   are as follows:
\begin{eqnarray}
T_s&=&\prod\limits^{j=N}_{j=0} T_{s{j+1\choose j}}\, , 
\quad \mbox{with}\nonumber\\[6pt]
T_{s{j+1\choose j}}&=&-\displaystyle{\frac 
{\left (Z_{s\,(j+1)}^\mathrm{into}+Z_{s\,(j+1)}\right )}
{\left (Z_{s\,(j+1)}^\mathrm{into}+Z_{s\,(j)}\right )}}
{e^{i\alpha_{j+1}}}
\label{Ts_into}\\[6pt]
\quad \mbox{and:}\nonumber\\[6pt]
T_p&=&\prod\limits^{j=N}_{j=0} T_{p\,{j+1\choose j}} \, , 
\quad \mbox{with}\nonumber\\[6pt]
T_{p\,{j+1\choose j}}&=&-\displaystyle{\frac {n_{j} Z_{p\,(j)}}{n_{j+1} {Z_{p\,{(j+1)}}}}}
\displaystyle{\frac {\left( Z_{p\,(j+1)}^\mathrm{into} + Z_{p\,(j+1)}\right)}{\left (Z_{p\,(j+1)}^\mathrm{into}+Z_{p\,(j)}\right )}}{e^{i\alpha_{j+1}}}
\label{Tp_into}
\end{eqnarray}
where ${T}_{j+1\choose j}$ are transmission coefficients at an interface between the $j$ layer and the $j+1$ layer.

\begin{figure}[h]
	\centering
\includegraphics[width=0.9\textwidth,keepaspectratio]{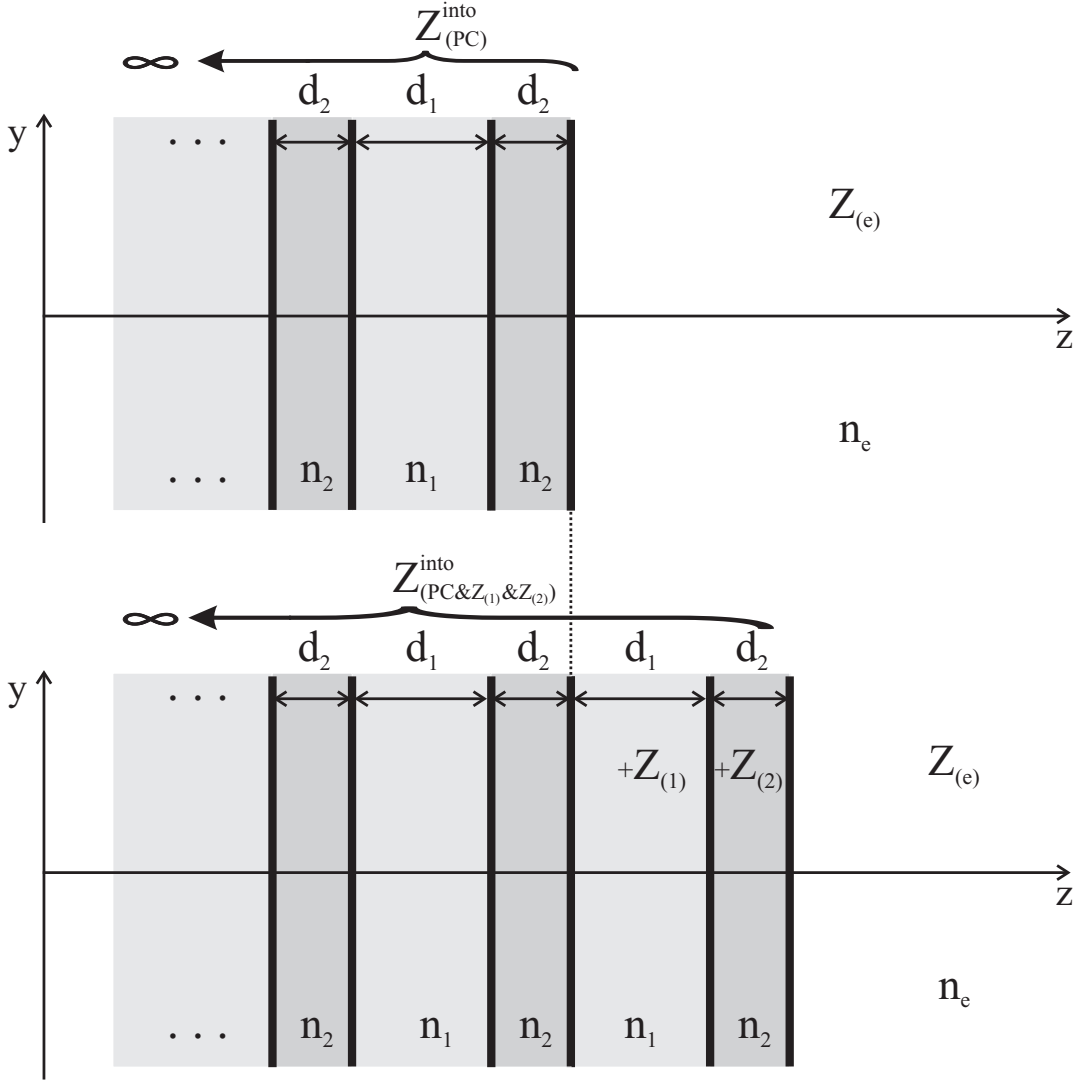}
		\caption{Determination of the impedance of the semi-infinite multilayer.
		  Note that $Z_{(PC\& Z_{(1)}\& Z_{(2)})}^\mathrm{into}\equiv Z^\mathrm{into}_{(PC)}$.}
	\label{Z_PC21}
\end{figure}

\subsection{The input impedance of a semi-infinite 1\mbox{-}D~PC}
Now, we are ready to obtain the input impedance of the 1\mbox{-}D~PC. 
 Let us assume that we have a  semi-infinite multilayer that consists of  alternative layers with impedances $Z_{(2)}$ and $Z_{(1)}$ and  its input impedance $Z^\mathrm{into}_{(PC)}$ is unknown (see the top of  Figure~(\ref{Z_PC21})). To find it, we make a next trick: first, we add an additional layer with impedance $Z_{(1)}$ to the multilayer and find the input impedance of this system using the recursion relation~(\ref{Z_into}):
\begin{equation}
Z_{(PC\& Z_{(1)})}^\mathrm{into}=Z_{(1)}{\frac {{ Z^\mathrm{into}_{(PC)}}-iZ_{(1)}\tan(\alpha_{1})}{Z_{(1)}-i{ Z^\mathrm{into}_{(PC)}}\tan(\alpha_{1})}}
\label{Z_PC_1}\, .
\end{equation}
Then, we add an additional layer with impedance $Z_{(2)}$ to this system and obtain the same semi-infinite multilayer again (see the bottom of Figure~(\ref{Z_PC21})) with
the input impedance:
\begin{equation}
Z_{(PC\& Z_{(1)}\& Z_{(2)})}^\mathrm{into}=
Z_{(2)}{\frac {{ Z^\mathrm{into}_{(PC\& Z_{(1)})}}-iZ_{(2)}\tan(\alpha_{2})}
{Z_{(2)}-i{ Z^\mathrm{into}_{(PC\& Z_{(1)})}}\tan(\alpha_{2})}}
\label{Z_PC_2_1}
\end{equation}
It is obvious that $Z_{(PC\& Z_{(1)}\& Z_{(2)})}^\mathrm{into}\equiv Z^\mathrm{into}_{(PC)}$, 
and, by solving equations (\ref{Z_PC_2_1}) and (\ref{Z_PC_1}) for  $Z^\mathrm{into}_{(PC)}$,  we obtain:
\begin{equation}
Z^\mathrm{into}_{(PC)}=-\frac{i}{2} {\frac{\left( (  Z_{(2)}^{2}-  Z_{(1)}^{2} )\tan
(\alpha_{2})\tan(\alpha_{1})\pm\sqrt {{s}}\right)}
{{  Z_{(2)}} \tan(\alpha_{1})+Z_{(1)}\tan(\alpha_{2})}}  ,
\label{Z_PC}
\end{equation}
where
$s=-4 {  Z_{(1)}} {  Z_{(2)}} \left ({  Z_{(2)}} \tan({  \alpha_1})+{  Z_{(1)}} 
\tan({  \alpha_2})\right )\left ({  Z_{(1)}} \tan({  \alpha_1})+{  Z_{(2)}}
 \tan({  \alpha_2})\right )+\left [\left (Z_{(2)}^2-Z_{(1)}^{2}
\right )\tan({  \alpha_1})\tan({  
\alpha_2})\right ]^2
$.

\subsection{Dispersion relation for surface waves}

Now, we find the dispersion relation for surface waves in 1-D structures ended by an arbitrary layer with impedance $Z_{(3)}$. 
(It may be a layer with any $n_3$,  for example a metal layer). 
The input impedance of such a structure will be:
\begin{equation}
Z_{(PC\& Z_{(3)})}^\mathrm{into}=Z_{(3)}{\frac {{ Z^\mathrm{into}_{(PC)}}-iZ_{(3)}\tan(\alpha_{3})}{Z_{(3)}-i{ Z^\mathrm{into}_{(PC)}}\tan(\alpha_{3})}}
\label{Z_PC_3}\, .
\end{equation}
A general condition for the existence of a surface wave between two media with impedances
$Z_{\mathrm{left}}$ and $Z_{\mathrm{right}}$ is:
\begin{equation}
Z_{\mathrm{left}}+Z_{\mathrm{right}}=0\, .
\end{equation}
In our case, this condition takes the form:
\begin{equation}
Z_{(PC\& Z_{(3)})}^\mathrm{into}+Z_{(e)}=0\, .
\label{SW_0}
\end{equation}
By solving equations~(\ref{SW_0}) and~(\ref{Z_PC_3}),  we
obtain the dispersion relation for the optical surface waves in the 1\mbox{-}D~PC:
\begin{equation}
\alpha_3\equiv\left[ k_{z(3)} d_3 \right] =  \pi  {M}+\arctan\left({\frac{-i\left({  Z^\mathrm{into}_{(PC)}}+{
  Z_{(e)}} \right){Z_{(3)}}}{Z_{(3)}^{2}+{  Z^\mathrm{into}_{(PC)}} {  Z_{(e)}}}}\right)  \, ,
\label{d_3} 
\end{equation}
where $M$ is a whole number.

This is a general dispersion relation, which is valid for both polarizations. 
If one would like to obtain 
the dispersion of the s-polarized optical surface wave, 
one should  use the $Z_s$ impedances in~(\ref{d_3}) and~(\ref{Z_PC}). 
Accordingly, to obtain the dispersion of the p-polarized optical surface wave, 
one must use $Z_p$ impedances in~(\ref{d_3}) and~(\ref{Z_PC}). 

It should be noted that the  solution obtained for s-polarization is equivalent 
(at $M=0$) to the
solution derived from a dispersion relation for an s-polarized optical surface wave,
which was deduced by Yeh, Yariv and  Hong using the unit cell matrix method 
(see equation~(58) in Ref.~\cite{Yariv1977}). 
 Meanwhile, the solution for p-polarization is equivalent to the solution
derived from the dispersion relation for a p-polarized optical surface wave,
which we deduced in a similar manner (using the unit cell matrix method) in Ref.~\cite{PRL2006}
(see equation~(2) in Ref.~\cite{PRL2006}). 
 The dispersion relation~(\ref{d_3}) and the relations presented in Refs.~\cite{Yariv1977, PRL2006}
were derived by different approaches 
(but, of course, both  started from the same Maxwell equations and boundary conditions)	 and presented in different terms, 
but these relations give the same results and may be transformed to each other.
 The advantages of the presented dispersion relation with the impedance terms  are its compact
structure and its unified form for both polarizations.
 In addition, a visible physical interpretation the impedance terms in the current dispersion 
relation allows it to be  easily extended for use with  more complicated structures. 
 For example, the addition of an adsorption layer between the layer~$_{(3)}$ and the external 
medium~$_{(e)}$
may be taken into account simply by changing of the impedance $Z_{(e)}$ by the impedance
$Z_{(e)}^\mathrm{into}$ calculated via recursion relation~(\ref{Z_into}),
where impedance of the external medium $Z_{(e)}$ is convoluted with an impedance of the adsorption layer $Z_{(a)}$.

\subsection{Band gap maximum extinction per length}
As a rule, in practical applications, we have  values  of two RIs of alternative media in 
the 1\mbox{-}D~PC, and the purpose is to find the thicknesses of  each alternative layer, 
which provides the maximum extinction per length at  given RIs, wavelength and angle. 
 Below, we derive this condition for the maximum extinction, 
which  makes it possible to minimize overall thickness of 1\mbox{-}D~PC.
 Also, we  show that 
the commonly-held opinion 
that it is  a ``quarter-wave-length''  thickness of the layers that provide the maximum extinction per length is incorrect.

To find the condition for the maximum extinction we derive the transmission coefficient for
one period of the PC (two layers), that is, the transmission through 
the three interfaces: 
\begin{equation}
T_{PC(j+3)\choose PC(j)} =
T_{PC(j+1)\choose PC(j)} T_{PC(j+2)\choose PC(j+1)} 
T_{PC(j+3)\choose PC(j+2)}\, , 
\end{equation}
where expressions for  $T_{j+1\choose j}$ are given by equations~(\ref{Ts_into}) 
or~(\ref{Tp_into}).
As a result we obtain for both polarizations:

\begin{equation}
T_{{PC(j+3)\choose PC(j)}} =\frac {\left( Z_{(2)}+ Z_{(PC)}^\mathrm{into}\right) \left(Z_{(1)}+ Z_{(PC\&Z_{(1)})}^\mathrm{into}\right)}
{\left(Z_{(1)}+ Z_{(PC)}^\mathrm{into}\right)
\left(Z_{(2)}+Z_{(PC\&Z_{(1)})}^\mathrm{into}\right )}\,e^{i(\alpha_1+\alpha_2)}\, , 
\label{T_123}
\end{equation}
where  $Z_{(PC\&Z_{(1)})}^\mathrm{into}$  is given by~(\ref{Z_PC_1}) 
and $Z_{(PC)}^\mathrm{into}$ is given by~(\ref{Z_PC}).


The desired values of the thicknesses 
$d_1=d_{1\mathrm{max}}$ and $d_2=d_{2\mathrm{max}}$ are the thicknesses at which  
the  next expression:
\begin{equation}
f(d_1,d_2)=
{\left|\frac {\ln \left(\left|T_{{PC(j+3)\choose PC(j)}}\right|\right)}{d_1+d_2}
\right|}
\label{max}
\end{equation}
reaches its maximum. 
 Absolute values are inserted in equation~(\ref{max})  in order to obtain the same
result regardless of the $\pm$ sign in equation~(\ref{Z_PC}).
 It may be noted that the ``quarter-wave-length''  thicknesses of the layers,
$d_j(\theta_j)=\lambda/(4n_j\cos(\theta_j))$, are values that
maximize another expression, namely,  expression~(\ref{max}) 
without the denominator, i.e., $(d_1+d_2)\times f(d_1,d_2)$  and 
these ``quarter-wave-length''  thicknesses are not optimal, especially 
at large incident (grazing) angles (i.e. at $\cos(\theta_j)\rightarrow 0$).

Now, we have  all the equations required to calculate a 1\mbox{-}D~PC structure for any particular 
experimental conditions.
 Presented below is an example in which we  obtain  the 1\mbox{-}D~PC structure ended by a Pd nanolayer,
which were successfully used for hydrogen detection in Refs.~\cite{OL2009},~\cite{NJP2009}.
 
 \section{Calculation of the 1\mbox{-}D~PC structure for particular experimental conditions}
 
\subsection{Angle-dependent variables }
It is very convenient to use a numerical aperture $\rho=n_0\sin(\theta_0)=k_y(\lambda/2\pi)$ 
as an angle variable instead of angles $\theta_j$ in each $j$-th layer and 
hereafter   we will  do so.
It is a unified angle variable for all layers, since according to Snell's law $\rho=n_0\sin(\theta_0)=n_j\sin(\theta_j)$, for any~$j$.
The angle-dependent variables have the next forms as functions of $\rho$:

\begin{eqnarray}
Z_{s(j)}=
&\displaystyle\frac{1}{n_j \cos(\theta_j)}
&=\displaystyle\frac{1}{n_j\sqrt{1-(\rho/n_j)^2}}   \qquad \mbox{\small(for TE wave)}
\label{Zs_rho}\\[8pt]
Z_{p(j)}=
&\displaystyle\frac{\cos(\theta_j)}{n_j}   
&=\displaystyle\frac{\sqrt{1-(\rho/n_j)^2}}{n_j} \qquad \mbox{\small(for TM wave)} 
\label{Zp_rho}\\[8pt]
k_{z(j)}=
&\displaystyle\frac{2\pi}{\lambda}n_j\cos(\theta_j)
&=\displaystyle\frac{2\pi}{\lambda}n_j\sqrt{1-(\rho/n_j)^2}   
\qquad \mbox{\small(for both polarizations)}\, .
\label{k_z}	
\end{eqnarray}

\subsection{Conditions for long-range propagation of the plasmon-polariton
waves in metal films}

A value of $\rho$ at which we would like to excite a PC~SW 
(and, therefore, a desired  SW wavevector  
$k_\mathrm{SW}=2\pi\rho/\lambda)$ depends on the particular problem
that we are trying to solve.
As one example, to excite LRSPPs in thin metal films, the effective RI of the LRSPPs 
$\rho$ should be close to the RI of the external medium $n_e$.
To prove this statement, in this subsection we will find the angular value $\rho$ at which the
electric field of the incident p-polarized wave has a minimum inside a thin layer
with a {\sl large extinction} for optical waves.
This minimum coincides with the zero of the main, tangential component $E_y$ of the
electric field in the film with the large extinction.
 This large extinction, i.e., large imaginary part of the RI of  a thin film material  
($\mathit{Im}(n_\mathrm{M})>>1$), may arise both  from a large negative permittivity 
of the  material ($\mathit{Re}(\varepsilon_\mathrm{M})<<-1$, 
even at small $\mathit{Im}(\varepsilon_\mathrm{M})$, as is the case for  silver or gold) and from  
a large losses in the  material ($\mathit{Im}(\varepsilon_\mathrm{M})>>1$, 
even at small or positive $\mathit{Re}(\varepsilon_\mathrm{M})$, as is the case for  
palladium).
 In both cases the optical wave does not penetrate deeply in the bulk material with permittivity
$\varepsilon_3=\varepsilon_\mathrm{M}$.
 It should be also noted that, in a very thin film,  additional  losses  always appear  
due to collision-induced scattering  of conducting electrons at the  walls of the  film.
 Therefore,  the imaginary part of the permittivity in  such a film 
is increased in comparison with the
bulk material (see Appendix~B for details).

So, let us assume that we have a thin film with RI $n_3=n_\mathrm{M}$ and that a p-polarized light wave is incident
on it with angular parameter $\rho$. 
The instantaneous value of the tangential component of the electric field 
at a coordinate point $z$ inside the film is the sum of the 
progressive 
and the recessive (reflected) waves:
\begin{eqnarray}
E_y(z)&=&E_{(+)} e^{-ik_{z(3)}(d_3-z)} + E_{(-)}  e^{ik_{z(3)}(d_3-z)} \nonumber\\
&=&E_{(+)} e^{-ik_{z(3)}(d_3-z)} + E_{(+)} R_{e\choose{^{_3}}}
 e^{ik_{z(3)}(d_3-z)}\, ,	
\label{E_y}
\end{eqnarray}
where the reflection coefficient from the interface between the film~$_{(3)}$  and
the external medium~$_{(e)}$ is given by~(\ref{R_into}): 
\begin{equation}
R_{e\choose{^{_3}}}= 
\displaystyle\frac{
{Z}_{(e)}-{Z}_{(3)}}{
{Z}_{(e)}+{Z}_{(3)}}\, . 
	\label{R_3_e}
\end{equation}

To find the coordinate $z_0$ at which the tangential component of the electric field
in the film is zero, we solve the equation: 
\begin{equation}
	E_y(z_0)=0
	\label{E_y_0}
\end{equation}
with  respect to $z_0$.  
Then we express the coordinate $z_0$ as
$z_0=d_3(1-\alpha)$, where
$\alpha$ is the coordinate
of the zero minimum $z_0$ in terms of  film thickness $d_3$.  
When $ z_0=0$ ($=> \alpha=1$), the zero  of the  $E_y$ takes 
place at an internal border of the film.  
When $ z_0=d_3/2$ ($=> \alpha=1/2$),  the field zero 
occurs in the center of the film.  
When $ z_0=d_3$ ($=> \alpha=0$),  the field zero  is located at an external
border of the film.
The solution of equations~(\ref{E_y_0},\,\ref{E_y}) is:
\begin{equation}
\alpha\equiv
\left[ {1-(z_0/d_3)} \right]=
-\displaystyle\frac{i\ln\left(-1/R_{e\choose{^{_3}}}\right)}{2k_{z(3)}d_3}\, .
	\label{alpha}
\end{equation}

Assuming that
$|n_{3}|>>\rho\geq n_{e}$ we 
will use the impedance of the final (metal) film in the so-called
Leontovich approximation (see, for example, Ref.~\cite{Senior1960} for details). 
In this approximation equations~(\ref{Zp_rho}) and~(\ref{k_z}) have the form:
\begin{eqnarray}
Z_{p(3)}=
&\displaystyle\frac{\sqrt{1-(\rho/n_3)^2}}{n_3}   
&\approx\displaystyle\frac{1}{n_3} 
\label{Zp_rho_L}\\[8pt]
k_{z(3)}=
&\displaystyle\frac{2\pi}{\lambda}n_3\sqrt{1-(\rho/n_3)^2}
&\approx\displaystyle\frac{2\pi}{\lambda}n_3\, ,
\label{k_z_L}
\end{eqnarray}
while 
\begin{equation}
	Z_{p(e)}=\displaystyle\frac{\sqrt{1-(\rho/n_e)^2}}{n_e}   
	\label{Zp_e}
\end{equation}
as usual.

Substituting~(\ref{Zp_rho_L}), (\ref{k_z_L}) and~(\ref{Zp_e}) in 
$R_{e\choose{^{_3}}}$~(\ref{R_3_e}) and
in  equation~(\ref{alpha}) and then solving it for $\rho$ we get (in the approximation $d_3<<\lambda$):
\begin{equation}
\rho_{\alpha}= 
n_e
+
2n_e^{3}\,\left[{\alpha}{\pi}{\frac {{d_3}}{{\lambda}}}\right]^{2}
\; ,
\label{rho1}
\end{equation}
which we
presented earlier   
(see eq.~(1) in Ref.~\cite{PRL2006}).

So, we have derived  equation~(\ref{rho1}), which  shows that   the
modulus of the electric field strength $|E_y|$  has a
minimum equal to zero when the p-polarized electromagnetic
wave incident on the thin (metal) film  in the angular range
from $\rho_{1}$ to $\rho_0$.
 In this case, the point at which $|E_y|=0$ is changed  from 
the internal ($\alpha=1, z_0=0$) to the
external ($\alpha=0, z_0=1$) borders  of the film,  and it is located in the center  of
the film at $\rho=\rho_{1/2}$ ($\alpha=1/2, z_0=1/2$).

Now the next question arises: are there some electromagnetic
surface modes with a wavevector in  the range: $k
=[\frac{\omega}{c}\rho_{1} \dots \frac{\omega}{c}\rho_0]$?
 If the answer is yes, one may expect that these modes will be long-range
propagated surface modes, since they may be excited by the
p-polarized waves incident on the film in the angular range $\rho
=[\rho_{1}\dots\rho_0]$, and these modes have the zero minimum
of  $|E_y|$ inside the lossy (metal) film.

One example of such a mode is widely recognized -- it is
 the LRSPP
in a thin film embedded between two identical dielectrics. 
 It is  known that the dispersion curve of  SPPs
splits as a result of a coupling between SPPs from both
film interfaces, and the LRSPP wavevector  shifts (at a
given frequency) to a light curve (i.e., $\rho\rightarrow n_e$). 
 The value of the LRSPP
wavevector is $k\simeq\frac{\omega}{c}\rho_{1/2}$ (see Ref.~\cite{Sambles1991} and/or Appendix~A),
where $\rho_{1/2}$ is given by~(\ref{rho1}) at $\alpha=1/2$, and the zero minimum of
$|E_y|$ is always located in the center of the film.

PC~SWs are 
another example of electromagnetic surface modes that
can propagate along thin metal films and  have a wavevector in the
range of $k =[\frac{\omega}{c}\rho_{1} \dots
\frac{\omega}{c}\rho_0]$. 
 In the next subsection we show how to design the PC structure for excitation PC~SW at 
desired $\rho\simeq[\rho_{1}\dots\rho_0]$ at given $\lambda$, $n_1$ and $n_2$.

\subsection{Derivation of $d_1$, $d_2$ and $d_3$ for PC structure terminated by Pd nanofilm}

\begin{figure}[h]
	{\centering
\includegraphics[width=\textwidth,keepaspectratio]{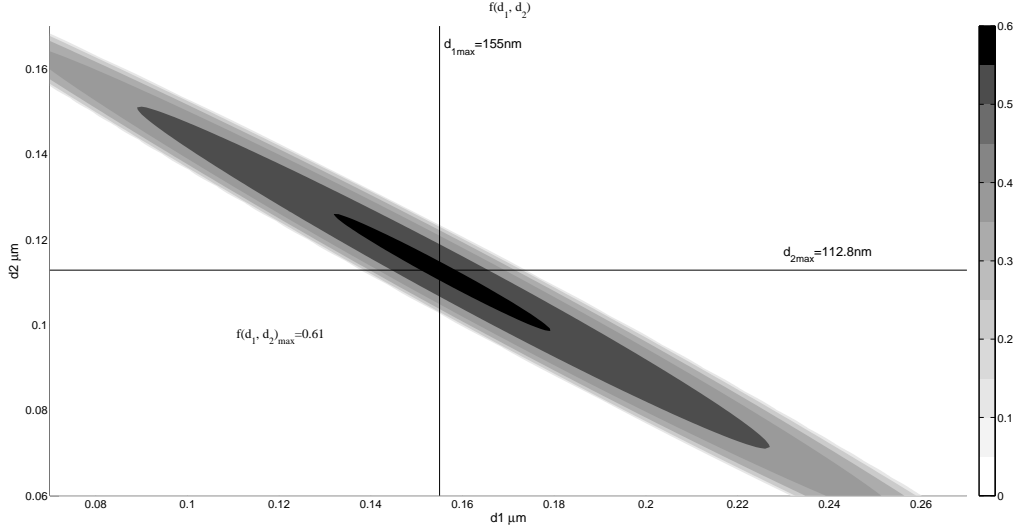}
}
		\caption{1\mbox{-}D~PC band gap extinction as a function of layers thicknesses.}
	\label{f_max}
\end{figure}

 So, let us to design PC structure with  a maximum band gap extinction at desired $\rho\simeq\rho_{1/2}\simeq 1.0012$ (obtained from (\ref{rho1}) at  $n_e=1.0003$, $d_3\simeq10$~nm and $\lambda\simeq 739$~nm). 
 Let us have two layer materials, $Ta_2O_5$ and $SiO_2$, with RIs of $n_2=2.076$ and
$n_1=1.455$ at the wavelength of our tunable diode laser  (at $\lambda\simeq 739$~nm).
 The function $f(d_1,d_2)$ given by~(\ref{max}) is presented in Fig.~\ref{f_max}.
 One can see that $f(d_1,d_2)$ reaches its maximum at 
$d_{1\mathrm{max}}=155.0$~nm and $d_{2\mathrm{max}}=112.8$~nm.
 At these thicknesses, the maximum  band gap extinction per length occurs, and,
therefore, the PC structure will have a minimal overall thickness. 

Now, we must find the thickness $d_3$ of the final palladium layer that will satisfy the dispersion relation~(\ref{d_3}). With $M=0$  we have:
\begin{equation}
d_3  = \frac{\lambda}{2\pi n_3\sqrt{1-(\rho/n_3)^2}} \arctan\left({\frac{-i\left({  Z^\mathrm{into}_{(PC)}}+{
  Z_{(e)}} \right){Z_{(3)}}}{Z_{(3)}^{2}+{  Z^\mathrm{into}_{(PC)}} {  Z_{(e)}}}}\right)  \, .
\label{d_30} 
\end{equation}
 Substituting $n_3=1.9+i4.8$  and the other values pointed in the present subsection into~(\ref{d_30}) (and into (\ref{Zp_rho}), (\ref{k_z}) and (\ref{Z_PC}) correspondingly) we obtain
$d_3=1.2$~nm.
 This  is a rather small value and it is better for the thickness of the palladium layer to be in the range of
$8\dots 10$~nm to be sure that  a continuous film we will obtain during deposition. 
 If we want to retain the maximum band gap extinction  at the desired $\rho_{1/2}$ and, therefore,
do not want to change the optimal values $d_{1\mathrm{max}}$ and $d_{2\mathrm{max}}$,
we may decrease the thickness of just the  $Ta_2O_5$ layer, which is contiguous with the $Pd$ layer.   The decrease in the  thickness  of the last $Ta_2O_5$ layer from $d_2=112.8$~nm to $d'_2=103.4$~nm permits
us to increase the thickness of the $Pd$ layer to the value  $d_3=8$~nm and satisfy the SW excitation
condition at the desired angle $\rho_{1/2}$ and the desired wavelength $\lambda$.

\begin{figure}[h]
	{\flushleft
\includegraphics[width=\textwidth,keepaspectratio]{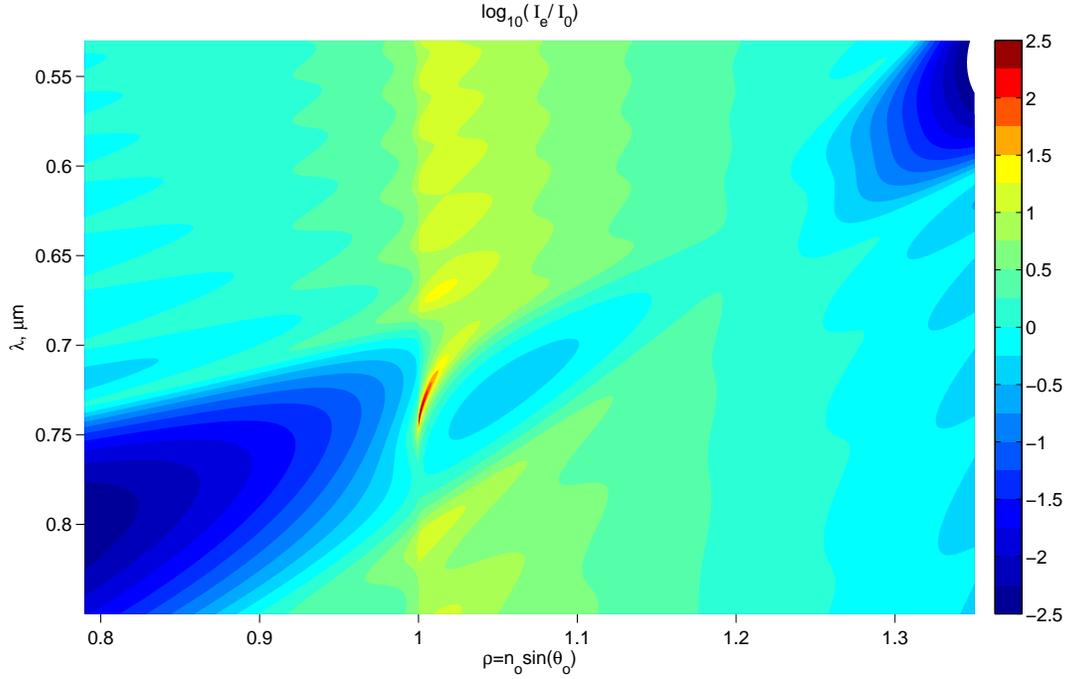}
}
		\caption{The calculated dispersion of the
1\mbox{-}D~PC structure with the terminal Pd nanolayer  in air.
The LRSPP  mode is  seen as the  red curve (with
an enhancement of more than 100) inside the band gap (blue areas with an enhancement of less
than 1). The
photonic band gap  vanishes near Brewster's angle
($\rho_{Br}\simeq 1.2$ in this system), where no reflection of the
TM wave takes place from  the $SiO_2/Ta_2O_5$ interface.}
	\label{disp737}
\end{figure}

Thus we derive the next PC structure:
\textit{substrate/$(HL)^{14}H'M$/air}, in which $H$ is a $Ta_2O_5$ layer
with a thickness $d_2=112.8$~nm, $L$ is a $SiO_2$ layer with
$d_1=155.0$~nm, $H'$ is a $Ta_2O_5$ layer with  $d_2'=103.4$~nm, and
$M$ is the palladium layer with  $d_3=d_M=8$~nm. 
 In  Fig.~\ref{disp737} the calculated dispersion  of
this 1\mbox{-}D~PC structure in air is presented as the logarithm of the
optical field enhancement
(i.e., as $\log T_{e\choose 0}$)
in the external medium near the structure.
 The optical field enhancement $T_{e\choose 0}$ is calculated by using (\ref{Tp_into}) 
and~(\ref{Z_into}) for 29 dielectric layers and one metal nanolayer on BK-7 ($n_0=1.513$) substrate.
 The optimal number of $Ta_2O_5/SiO_2$ pairs depends on a total extinction in the layers.
 In the present case, the $Pd$ nanolayer is the layer that gives the main contribution in extinction and the 14
pairs provide the optimal coupling (i.e., $R_{PC}=0$) of the  incoming EM radiation into SWs.
 
 The dispersion is presented using the coordinate $\lambda(\rho)$.
 The angular parameter $\rho$, at which the excitation of the surface mode occurs,  
is equal to the effective RI of the mode, $n_{\textrm{\scriptsize sw}}$.
 Therefore, the red curve that is inside the blue band gap in  Fig.~\ref{disp737} presents the dispersion of the
LRSPP mode (i.e., the dependence of its effective RI on
the wavelength). 
One can see from  Fig.~\ref{disp737} that in accordance with~(\ref{rho1})  the LRSPP mode exists  
near $\rho\sim n_e$ only
($\rho\simeq1\dots1.0026$), where the minimum of EM field inside the metal nanolayer 
occurs, while,
at $\rho>1.0026$ the  EM damping is increased significantly in this 8-nm thick metal film.

In the presented case
the optical surface mode dispersion
curve approaches the line of the total internal
reflection (TIR) ($\rho_\mathrm{TIR}\equiv n_e=1.0003$) at wavelengths in
the range of $\lambda\sim734\dots746$~nm. 
 Therefore, at these
wavelengths, we can excite the SPP at $\rho\rightarrow n_e$  and
expect  it to be LRSPP. 
 At $\lambda=733.7$~nm ($\rho=1.0026\simeq\rho_1$), the zero
minimum of $|E_y|$ occurs at the internal interface of the 8~nm palladium film,
while at $\lambda=740.2$~nm ($\rho=1.00088\simeq\rho_{1/2}$)
the zero
minimum of $|E_y|$ occurs in the center of this 8-nm thick nanofilm.
 Moreover, at $\lambda=745.6$~nm 
($\rho=1.000301$, i.e., $\rho\rightarrow\rho_0\simeq n_e$)  we can excite LRSPP with $|E_y|=0$
near the external  interface of the film, and  this wave will have the largest
propagation length. 
 Theoretical propagation lengths, i.e., propagation lengths that take into account the internal damping
in the Pd nanofilm but do not take into account the surface scattering of LRSPPs to photons, for these wavelengths are: $L_{733.7}=0.14$~mm, $L_{740.2}=0.32$~mm and $L_{745.6}=5.4$~mm.
 Therefore, contrary to
intuitive expectations, modes with $\rho= \rho_0\rightarrow
n_e$ 
($|E_y|=0$ at the external  interface of the film, i.e., modes with an asymmetric field distribution in the nanofilm) 
are more long-range propagated than the mode with
$\rho=\rho_{1/2}$ 
($|E_y|=0$ in the center of the film). 
More details about this issue are provided  in  Appendix~A.

\section{Conclusions}
The present article provides the theoretical background and the algorithm for the design of 1\mbox{-}D~PC structures that support the propagation of optical surface waves. 
 We have used the impedance approach, which permits calculation for s- and p-polarizations by the same equations.  
 The main results of this work are the input impedance of semi-infinite 1\mbox{-}D~PC~(\ref{Z_PC}) and the dispersion relation of PC~SWs~(\ref{d_3}). 
 Equations~(\ref{T_123}) and~(\ref{max})  are needed for
the design of a multilayer structure with  maximal band gap extinction at the  minimal overall thickness of the structure. 
 Equation~(\ref{rho1}) is important for understanding of the physical reasons for 
the    propagation of the LRSPP even in thin films with large extinction, while the equations
in  Appendix~A provide additional insight  concerning why  slightly asymmetrical structures provide
more long-range propagation of the surface waves.

\section*{Acknowledgments}
This work was  supported financially by the Russian Federal Program  ``Research and educational personnel of innovative Russia'' and by the Russian Foundation for Fundamental Research.
The author thanks E.V.~Alieva for useful discussions and assistance.

\newpage

\section*{Appendix A: Dispersion relation for LRSPPs in the symmetrical configuration and
increase of  propagation length associated with slight asymmetry in the system.}

\setcounter{equation}{0}
\renewcommand{\theequation}{A{\arabic{equation}}}

The propagation of LRSPP in a symmetrical structure (metal film between identical dielectrics)
has been considered in several publications~\cite{Sarid1981, Sambles1991}. 
Here we will consider this point based on the impedance approach and then will add slight asymmetry to the structure.  

The dispersion relation, $\lambda(\rho)$, for surface waves in a metal film between two semi-infinite dielectric media can be obtained easily, for example, from equation~(\ref{d_3}) or equation~(\ref{d_30}), 
in which the 1\mbox{-}D~PC is replaced by a uniform dielectric medium with a RI $n_0$ ($Z^\mathrm{into}_{(PC)}\rightarrow Z_{(0)}$): 
\begin{equation}
 \frac{1}{\lambda}={\frac{1}{2\pi d_3 n_3 \sqrt{1-(\rho/n_3)^2}}}\arctan\left({\frac{-i\left({  Z_{(0)}}+{
  Z_{(e)}} \right){Z_{(3)}}}{Z_{(3)}^{2}+{  Z_{(0)}} {  Z_{(e)}}}}\right)  \, .
\label{lambda_rho_SPP} 
\end{equation}

\begin{figure}[h]
	{\flushleft
\includegraphics[width=0.9\textwidth,keepaspectratio]{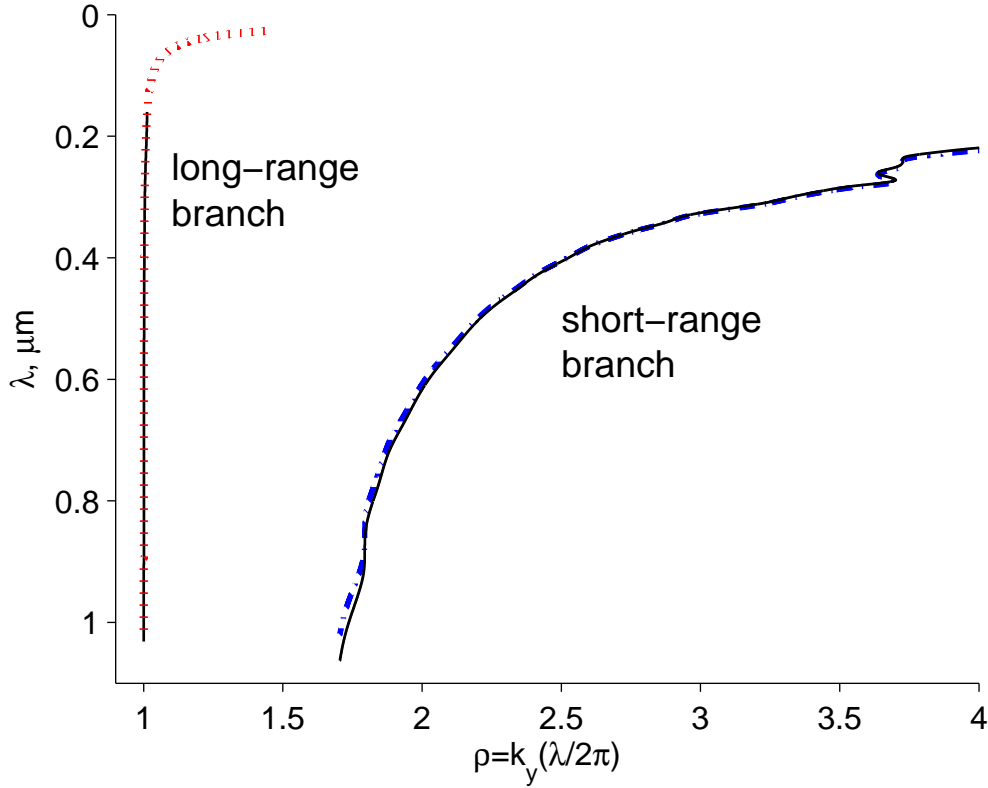}
}
		\caption{ The dispersion of a Pd nanofilm that is
		 freely suspended in the air.
The  solid  black curve is calculated  from the general dispersion relation 
for the symmetric system~(\ref{lambda_rho_symm}), 
while the dotted red curve and the	dashed-dotted blue curve are calculated by approximated
solutions~(\ref{rho_LRSPP}) and~(\ref{rho_SRSPP}), respectively.}
	\label{disp_symm}
	\end{figure}

For a pure symmetrical configuration [$n_0=n_e$ and $Z_{(0)}=Z_{(e)}$], the general dispersion relation~(\ref{lambda_rho_SPP}) has the  form shown below:
\begin{equation}
 \frac{1}{\lambda}={\frac{1}{2\pi d_3 n_3 \sqrt{1-(\rho/n_3)^2}}}\arctan\left({\frac{-2i  
  Z_{(e)}{Z_{(3)}}}{Z_{(3)}^{2}+ {  Z_{(e)}^2}}}\right)  \, 
\label{lambda_rho_symm} \, .
\end{equation}
By  taking into account the Leontovich approximation, (\ref{Zp_rho_L}) and (\ref{k_z_L}),
the equation
can be  simplified further:
\begin{equation}
 \frac{1}{\lambda}\simeq{\frac{1}{2\pi d_3 n_3}} \arctan\left(
 {\frac {{2\, n_3}\,{{n_e}}^{2}
 \left({\rho}^{2}-{{n_e}}^{2}\right)^{-1/2}}
{{{n_e}}^{4}+{{n_3}}^{2}{{n_e}}^{2}-{{n_3}}^{2}{\rho}^{2}}
}
 \right)  \, .
\label{lambda_rho_Leon} 
\end{equation}
For a very thin film (at $d_3\rightarrow 0$) one can obtain two solutions for $\rho$ 
from (\ref{lambda_rho_Leon}):
\begin{eqnarray}
\rho_\mathrm{LRSPP}&\simeq& 
n_e
+
\frac{n_e^{3}}{2}\,\left[{\pi}{\frac {{d_3}}{{\lambda}}}\right]^{2}
\label{rho_LRSPP}\\[8pt]
\rho_\mathrm{SRSPP}&\simeq&
{\frac {\pi{d_3}n_e^{2}}{3\lambda}}
-{\frac {\lambda n_e^{2}}{\pi \mathit{Re}(n_3^2){d_3}}}
-{\frac {\pi d_3 \mathit{Re}(n_3^2)}{2\lambda}}
\, .
\label{rho_SRSPP}
\end{eqnarray}
The former equation presents an approximation for the  long-range branch of the dispersion relation
of a symmetrical system (\ref{lambda_rho_symm}), 
while the latter is  an approximation for the  short-range branch (its damping increases at $d_3\rightarrow 0$). 
 In Fig.~{\ref{disp_symm}}, 
graphical illustrations of  dispersion relation~(\ref{lambda_rho_symm}) [or~(\ref{lambda_rho_Leon}),
since no  difference can be   seen  between them at this scale] and its 
approximations~(\ref{rho_LRSPP}) and~(\ref{rho_SRSPP}) are presented. 
 Calculations are performed  for
an  8-nm thick Pd film that is  suspended freely in the air ($n_e=1.0003$), and the  Pd RIs at different wavelengths  were taken from Ref.~\cite{Palik1985}.
  In addition to  good agreement between the dispersion equation for the symmetrical system and the approximated equations, (\ref{rho_LRSPP}) and~(\ref{rho_SRSPP}), Fig.~{\ref{disp_symm}} shows that the long-range branch of the curve has no noticeable wavelength dispersion in the visible range, while  the system with PC from one side the LRSPP curve  (see Fig.~\ref{disp737}) has  prominent dispersion.

It is apparent
that~(\ref{rho_LRSPP}) consides exactly  with~(\ref{rho1}) at 
$\alpha=1/2$, and, therefore, in a pure symmetrical  configuration,
the zero  of the tangential component and the total minimum  of the electric field 
are always located in the center of the metal nanofilm, which is not surprising due to the symmetry of the system.
 A more surprising fact is that this symmetrical system (with $|E_y|=0$ in the center of the film)  is not the optimal solution if one is looking for the smallest damping and the longest propagation length of the LRSPP.
 To the best of our knowledge 
Wendler and Haupt were the first to point out on this peculiarity. 
 In the theoretical article~\cite{Wendler1986} they  showed, through numerical calculations, 
that, 
in a slightly asymmetrical structure, the propagation length may be three orders of magnitude
greater than the propagation length in the symmetrical structure. 
 
 In experimental practice~\cite{PRL2006}, \cite{OL2009} we have been able to increase the 
propagation length of SPPs by a factor in the range of 100-200 times.
Further increase is limited by  damping due to  surface scattering, which 
occurs due to
the close proximity of the effective RI of the LRSPP to the RI of the external medium, 
or, in other words, ``the plasmon curve becomes too close to the light line''.
 Therefore,  any small disturbance can transform plasmons to photons when this zero minimum is  approaching  the border of the film (and, therefore, $\rho\rightarrow n_e$).
 But, nevertheless, the internal damping of LRSPPs can be decreased when slight asymmetry 
exists in the structure.

Below, we obtain analytical expressions for the optimal $\Delta n_\mathrm{max}$ difference between the RIs of the dielectrics and show that the propagation length of the LRSPP increases when the minimum of the electric field shifts to the interface of the film (but does not leave the borders of the film). 
 To do this, we  return to the general dispersion relation for a non-symmetrical system~(\ref{lambda_rho_SPP}) and assume that the  RIs of the  dielectrics on both sides  of the metal film differ  only by a small value $\Delta n$: 
\begin{equation}
n_0=n_e+\Delta n \, .
\label{Delta_n } 
\end{equation}
We are solving for  the LRSPP wave only, so $\rho$ will differ from $n_0$ by
a small value $\Delta\rho$:
\begin{equation}
\rho=n_0+\Delta\rho=n_e+\Delta n +\Delta\rho \, ,
\label{Delta_rho } 
\end{equation}
where $\Delta\rho\sim O(d_3^2/\lambda^2)$ -- see (\ref{rho_LRSPP}).

In the limit, with small $\Delta\rho$ and small $\Delta n$, the dispersion relation~(\ref{lambda_rho_SPP}) can be 
simplified to the form:
\begin{equation}
 \lambda^{-1}(\Delta\rho, \Delta n ) \approx \frac{1}{\pi d_3}\left(\frac{2}{n_e^9\,\Delta\rho }\right)^{1/2}
 \left[
 n_e^3\,\left(\Delta\rho+\frac{1}{4}\Delta n\right)
 -\left(\frac{1}{2}\,n_3^2 
 +\frac{9}{16}\,n_e^2 \right) \Delta\rho\, \Delta n
 \right]
 \, .
\label{LRSPP_n} 
\end{equation}
Solving this equation for $\Delta\rho$ and taking its imaginary part  we obtain the next extinction  of LRSPPs in a slightly asymmetrical system:

\vspace{-3ex}
{\samepage
$$
\mathit{Im}\left[\rho_\mathrm{LRSPP}(\Delta n)\right]
	=
	\mathit{Im}\left[n_0+\Delta\rho\right]\approx  
$$
	\begin{equation}
	\frac {\Delta n \mathit{Im}(n_3^2)}{32 n_e^3\lambda^2}
\left(16 \pi^2 d_3^2 n_e^3 +3\pi^2 d_3^2 \Delta n 
	\left(8 \mathit{Re}(n_3^2)+9 n_e^2\right)-8 \Delta n \lambda^2
\right)
\, .
\label{rho_LRSPP_n}
\end{equation}
}
This over-approximated imaginary part of the effective RI of the LRSPP has {\it two}
solutions with (approximately) zero extinction, 
$\mathit{Im}\left[\rho_\mathrm{LRSPP}(\Delta n)\right]=0$, and, therefore, 
with maximal propagation length, namely, the first solution is
$\Delta n=0$ and the second solution is:
\begin{equation}
\Delta n_\mathrm{max}\approx	16 \frac{\pi^2 d_3^2 n_e^3}{8 \lambda^2
-3 \pi^2 d_3^2 \left (8 \mathit{Re}(n_3^2) + 9 n_e^2
\right )}
\; ,
\label{n_max}
\end{equation}
which in the limit $d_3\approx 0$ is:
\begin{equation}
\Delta n_\mathrm{max0}\approx 
2n_e^{3}\,\left[{\pi}{\frac {{d_3}}{{\lambda}}}\right]^{2}
\label{n_max0}
\end{equation}
One can see that (\ref{n_max0}) coincides with the addition to $n_e$ from (\ref{rho1}) at $\alpha=1$
(i.e., when the minimum of the field is at the nanofilm border).
 Therefore, expression (\ref{n_max}) also may be  considered as a cutoff condition for the existence
of  LRSPPs (at $\Delta n>\Delta n_\mathrm{max}$ no bounded modes exist). 
 From (\ref{lambda_rho_SPP}), it can be determined that,  at 
$n_e= n_0-\Delta n_\mathrm{max}$, the effective RI of LRSPP is infinitesimally close to the TIR angle,
i.e., $\rho\rightarrow n_0$. 

It is worth  noting here that the  analytical expression for $\Delta n_\mathrm{max}$ from (\ref{n_max}) corresponds well with
numerical data presented by 
Wendler and Haupt in Ref.~\cite{Wendler1986}. 
 Even the over-approximated expression for $\Delta n_\mathrm{max0}$ from (\ref{n_max0}), which is not dependent on the nanofilm RI $n_3$, 
is coincident with these numerical data at small film thicknesses (i.e.,  $d_3\leq 15$~nm).

At excitation of ultra-long-range SPP in thin films by introducing a small asymmetry in the system 
and, therefore, by abutting the plasmon curve on the light line, it is very important to maintain 
an appropriative  
balance between decreasing  the internal damping of LRSPPs and   increasing the scattering  of 
the LRSPPs.
The possibility for tuning   system parameters during the experiment to find the optimum is 
extremely
convenient in this case.

The principal difference between \textit{D/M/D'} and  \textit{D/M/PC} systems is that, in  the  system with a 1\mbox{-}D~PC on  one side, it is experimentally easy to introduce such a small asymmetry (in effective RI) by tuning the wavelength of the laser,
while it is really difficult to (finely) tune the dielectric constants  on the one interface by any means. 
 Wavelength tuning would be not useful in the \textit{D/M/D'} case, because  dispersion of dielectrics is small in non-absorbing wavelength regions, 
while  wavelength tuning in
the  PC structure is very effective due to  the high wavelength dispersion  inside the band gap regions.

\newpage

\section*{Appendix B: Additional damping in thin film.}

\setcounter{equation}{0}
\renewcommand{\theequation}{B{\arabic{equation}}} 

Damping of EM waves in thin metal films increases due to collision-induced scattering  of conducting electrons at the walls of the film.  
 This occurs when the mean free path of the electrons becomes comparable to the characteristic dimensions of the system.
 Therefore,  the imaginary part of the permittivity in  such a film 
is increased in comparison with the
bulk material.
 Unfortunately in many articles (especially theoretical articles), authors have usually disregarded this fact 
and used overly optimistic values for the  imaginary part of the metal permittivity.

Here we present  a summary of  formulae for this case, which help to estimate the minimal addition
to  $\mathit{Im}(\varepsilon_\mathrm{M})\equiv\varepsilon_\mathrm{bulk}''$.
 At optical frequencies, the damping in  nanostructures 
with the characteristic dimension $L$ is
 \begin{equation}
\gamma=\gamma_\mathrm{bulk}+\frac{v_F}{L}\, ,
\label{Bohren}
\end{equation}
where  $\gamma_\mathrm{bulk}$ is the damping constant for the bulk sample and 
$v_F$ is the electron velocity on the Fermy surface.
Below,  several specific cases are considered.

\subsubsection{Nanofilm as a set of nanoparticles:}
\noindent For sphere~\cite{Bohren}
 \begin{equation}
\varepsilon''
= 
\frac{\omega_p^2}{\omega^3}\gamma
=
\frac{\omega_p^2}{\omega^3}\left (\gamma_\mathrm{bulk}+ \frac{3 v_F}{4R}\right )
= 
\varepsilon_\mathrm{bulk}''+ 
\frac{3}{4}\frac{\omega_p^2}{\omega^3}\frac{ v_F}{R}\, ,
\label{Bohren2}
\end{equation}
where $\omega_p$ is plasma frequency. Therefore,
the characteristic dimension in this case  is
$L=L_R=4R/3$  ($R$ is the radius of the sphere).

\subsubsection{Continuous film:}
\noindent For continuous film with thickness $d$~\cite{Theye}:
 \begin{equation}
\gamma=\gamma_\mathrm{bulk}+
 \frac{3\omega_p}{8}
\frac{v_F}{c}
\frac{1+\cosh^2({\omega_p d}/{c})}
{\sinh({\omega_p d}/{c})\cosh({\omega_p d}/{c})+
{\omega_p d}/{c}}
\, .
\label{Theye1}
\end{equation}
For  very thin film (nanofilm),
in the limit $d\rightarrow 0$ 
 \begin{equation}
\gamma=\gamma_\mathrm{bulk}+
 \frac{3}{8}
\frac{v_F}{d}
\, ,
\label{Theye01}
\end{equation}
(therefore $L=L_d=8d/3$ in this limit) and 
 \begin{equation}
\varepsilon''
= 
\frac{\omega_p^2}{\omega^3}\left (\gamma_\mathrm{bulk}+ \frac{3 v_F}{8d}\right )
= 
\varepsilon_\mathrm{bulk}''+ 
\frac{3}{8}\frac{\omega_p^2}{\omega^3}\frac{ v_F}{d}\, ,
\label{Theye2}
\end{equation}
while
in the general case 
 \begin{equation}
\varepsilon''
= 
\varepsilon_\mathrm{bulk}''+ 
 \frac{3\omega_p^3}{8\omega^3}
\frac{v_F}{c}
\frac{1+\cosh^2({\omega_p d}/{c})}
{\sinh({\omega_p d}/{c})\cosh({\omega_p d}/{c})+
{\omega_p d}/{c}}
\, .
\label{Theye2g}
\end{equation}

 We have considered only collision-induced  damping,  and we did not take into account other types of 
damping in small volumes, which may lead to further increases of the imaginary part of the nanosized materials at optical frequencies. 
 
The collision-induced damping increase in the nanofilm at zero frequency may be estimated   by Fuchs' formula~\cite{Fuchs-Larson1971} for  a specific electrical
resistance of continuous films. 
For example, in our work with gold nanofilm~\cite{PRL2006},
the electrical resistance of the nanofilm was 850$\Omega$ for a strip that was with  0.25 mm wide  and  6.5 mm long. Therefore, for our nanofilm thickness of 5~nm, the specific resistance in the nanofilm was $\rho_\mathrm{film}=16.3\,\mu\Omega$cm,
which is 7.4 times greater than the specific
resistance  of  bulk gold ($2.2\,\mu\Omega$cm).
 While according to Fuchs formula at $d<<l_0$~\cite{Fuchs-Larson1971}
  \begin{equation}
\frac{\rho_\mathrm{bulk}}{\rho_\mathrm{film}}=\frac{3d}{4l_0}\ln\left(\frac{l_0}{d}\right)
\; .
\label{Fuchs-Larson}
\end{equation}
Substituting
the mean free path of electrons in bulk gold, which at room temperature is $l_0\simeq 50$\,nm,
we obtain that the specific
resistance of the 5~nm gold film at room temperature should be
about six times greater than that of bulk gold. This value is in
good agreement with the measured value, taking into account
imperfections
of the sputtered gold nanofilm.


\end{document}